# The mechanics of rocking stones: equilibria on separated scales

**Gábor Domokos, András Árpád Sipos and Tímea Szabó**

Department of Mechanics, Materials and Structures, Budapest University of Technology and Economics, Hungary, e-mail: domokos@iit.bme.hu

## Abstract

Rocking stones, balanced in counter-intuitive positions have always intrigued geologists. In our paper we explain this phenomenon based on high-precision scans of pebbles which exhibit similar behavior. We construct their convex hull and the heteroclinic graph carrying their equilibrium points. By systematic simplification of the arising Morse-Smale complex in a one-parameter process we show that equilibria occur typically in highly localized groups (flocks), the number of the latter can be reliably observed and determined by hand experiments. Both local and global (micro and macro) equilibria can be either stable or unstable. Most commonly, rocks and pebbles are balanced on stable local equilibria belonging to stable flocks. However, it is possible to balance a convex body on a stable local equilibrium belonging to an unstable flock and this is the intriguing mechanical scenario corresponding to rocking stones. Since outside observers can only reliably perceive flocks, the last described situation will appear counter-intuitive. Comparison of computer experiments to hand experiments reveals that the latter are consistent, i.e. the flocks can be reliably counted and the pebble classification system proposed in our previous work (Domokos et al 2010) is robustly applicable. We also find an interesting logarithmic relationship between the Zingg parameters and the average number of global equilibrium points, indicating a close relationship between the two systems.

## 1. Introduction

### 1.1. Pebble surfaces and rocking equilibria

Rocking stones, i.e. large rocks finely balanced in 'unlikely' positions have not only posed a puzzle for geologists (Linton 1955) but also were associated with witchcraft (Wikipedia: Rocking Stones) and presented a challenge for artists (Wikipedia: Rock balancing). These stones can be rocked by small effort despite their large weight. There is considerable interest in using rocking stones to determine the earthquake history of a given region (Bell et al 1998, Anooshehpoor et al 2004, Brune et al 2006). In this paper we propose a simple mechanical explanation for the existence of rocking stones by taking a close look at pebble and rock morphology. While pebble and surfaces are not convex, pebbles roll along the convex hull of their surface and this hull can be best approximated by a many-faceted polyhedron (Fig. 1).

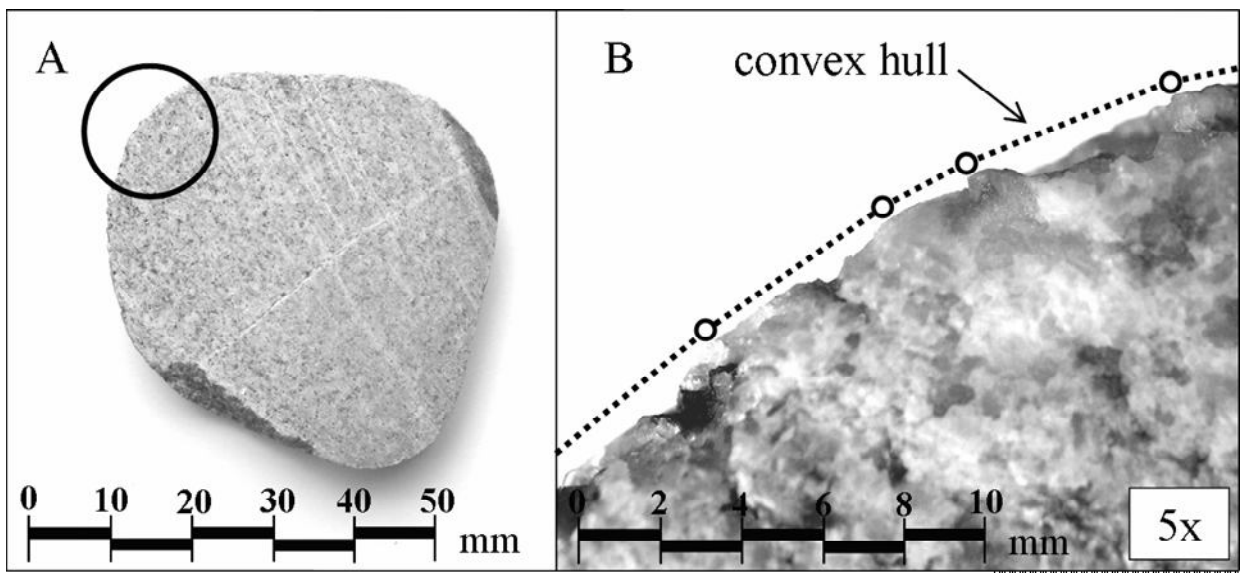

**Figure 1**: A, Planar cross section of pebble B, Magnified planar section and approximate convex hull

Our goal is to show that on this type of surface equilibrium positions exist typically on two scales. Smooth, convex curves and surfaces typically exhibit isolated equilibrium points (Fig. 2A). By approximating the smooth surface (or curve) by a discrete polyhedron (polygon), the number of equilibria often increases, however, the additional equilibrium points are not uniformly scattered on the surface. Rather, they accumulate in *flocks,* centered around the equilibria of the original smooth surface (Fig. 2B).

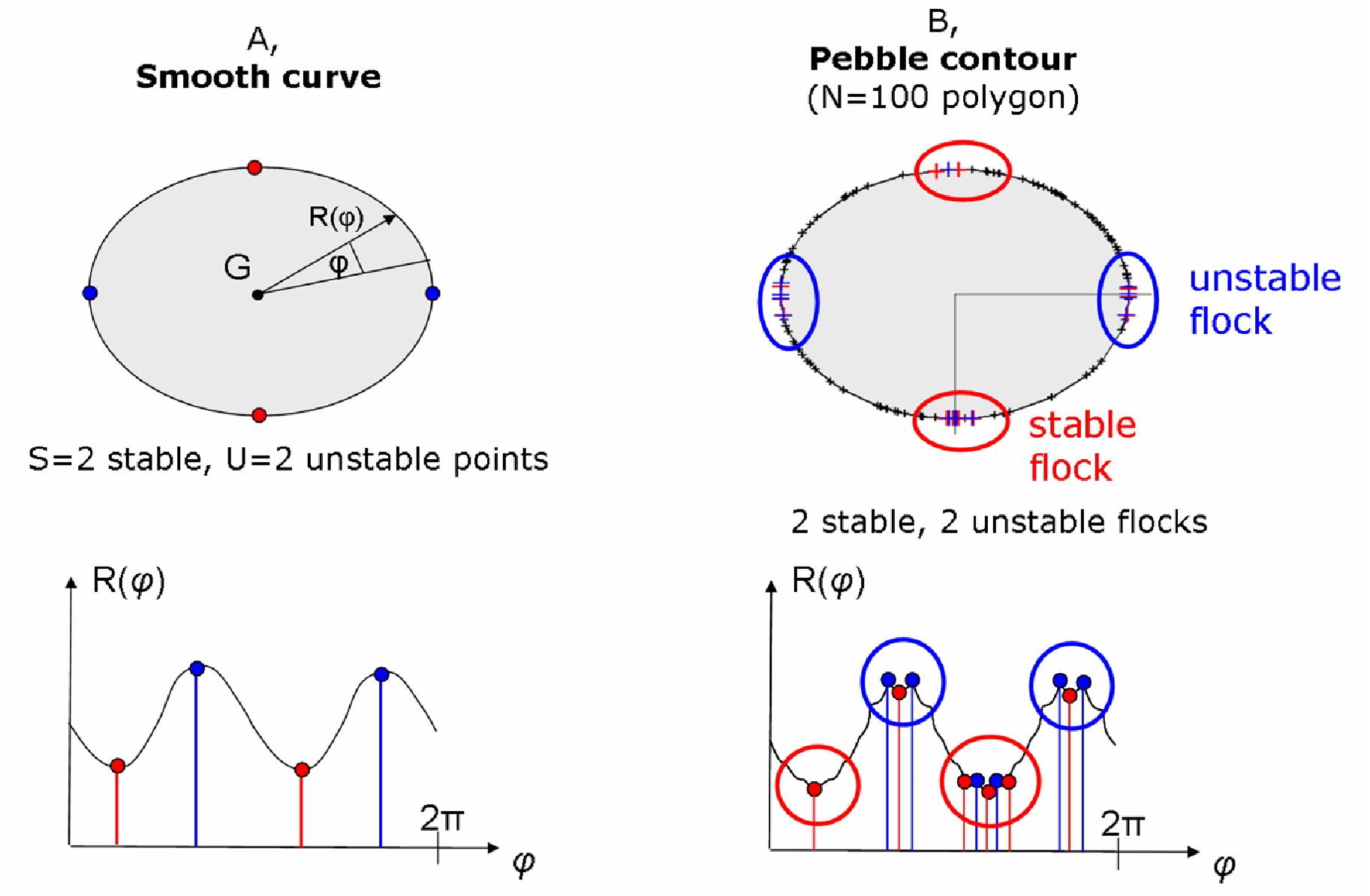

**Figure 2**: Equilibria of planar curves. Curve is characterized as scalar R distance from center of gravity G. Red/ blue points denote stable/ unstable equilibria, respectively. A, Smooth ellipse with S=2 stable and U=2 unstable points. B, N=100 discretization with S=9 stable and S=9 unstable points, accumulated in 2 stable and 2 unstable flocks.

We can also describe the stability properties of flocks. By characterizing the curve (or surface) with the scalar distance R from the center of gravity G as a function of the polar angle φ, (in 3D we have two angles φ, θ) stable micro-equilibria correspond to local minima, unstable micro-equilibria to local maxima of R. *Stable flocks* correspond to a set of nearby micro-equilibria located in a large potential valley, separated by small energy thresholds, *unstable flocks* correspond to similar collection of micro-equilibria in the vicinity of a large potential peak (or saddle). Flocks may exhibit various topologies, some 2D and 3D examples are illustrated in Figure 3.

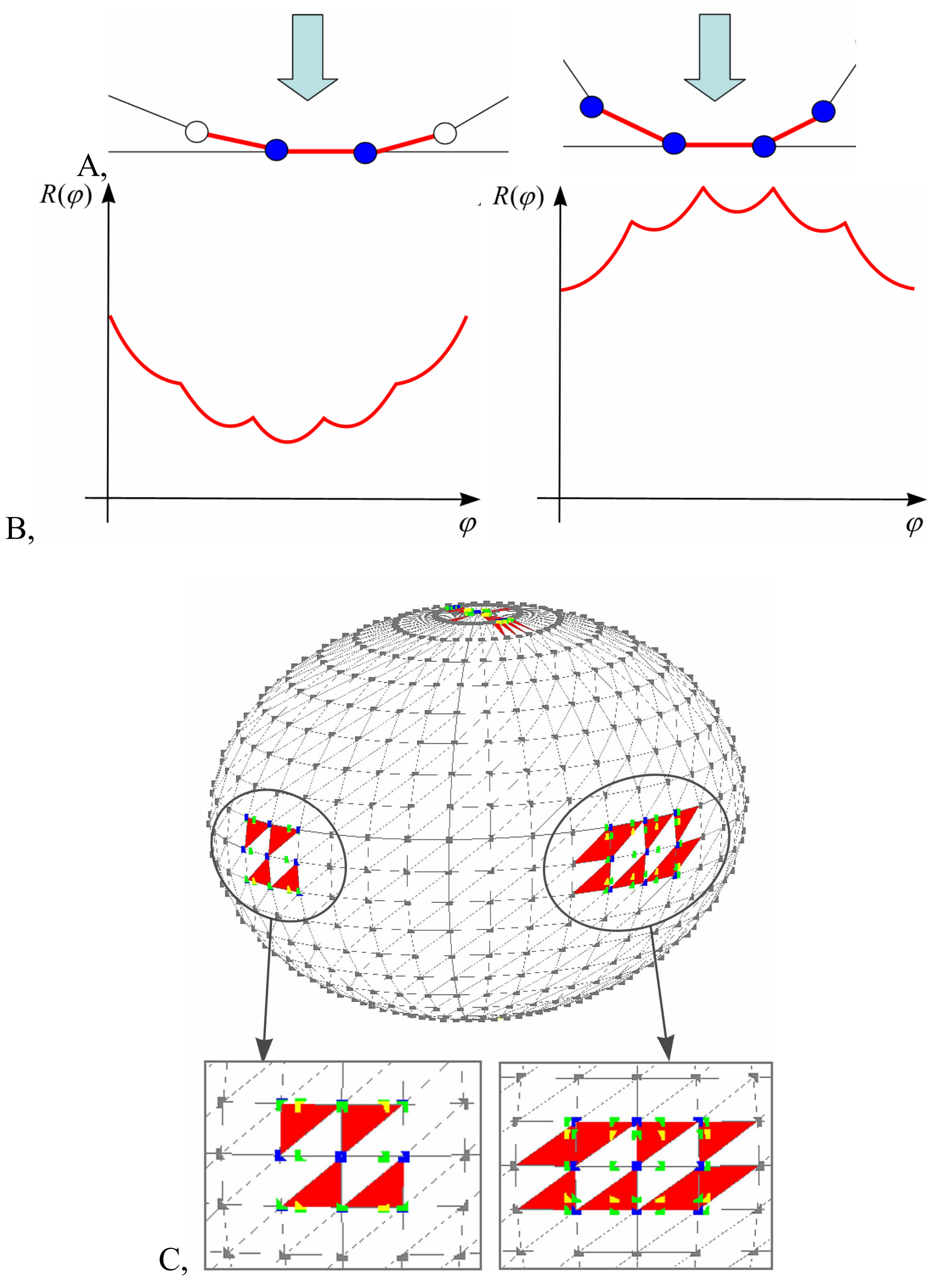

**Figure 3**: Illustration of flocks A, stable and unstable 2D flocks appearing on the physical contour B, stable and unstable 2D flocks appearing in the potential R(φ) C, some 3D flocks. Red faces with yellow points denote stable equilibria, unstable/ saddle points are marked with blue/ green, respectively.

Somewhat surprisingly, flocks can be present in *arbitrarily fine* discretizations. In case of the planar ellipse the computation is quite straightforward. By uniformly discretizing (according to arclength) an ellipse with S=2 stable and U=2 unstable equilibrium points and with axes ratio $1{:}a$ ($a \geq 1$) into a polygon with N=4k-2 sides and let k →∞, we can observe in Figure 4 that the four flocks (two stable and two unstable) shrink in diameter, however, the number of contained micro-equilibria approaches a non-trivial constant which only depends on the axis ratio *a*. Table 1 shows the numbers of stable and unstable micro-equilibria in the k →∞ limit, contained in stable and unstable flocks, respectively.

| *Type of flock ->* <br> *Type of micro-equilibrium* | *Stable* | *Unstable* |
|---|---|---|
| *Stable* | $2\left\lfloor \frac{a^2}{2(a^2-1)} \right\rfloor + 1$ | $2\left\lfloor \frac{a^2}{2(a^2-1)} \right\rfloor$ |
| *Unstable* | $2\left\lfloor \frac{a^2}{2(a^2-1)} \right\rfloor$ | $2\left\lfloor \frac{a^2}{2(a^2-1)} \right\rfloor + 1$ |

**Table 1**: Number of micro-equilibria on the ellipse with infinitely fine, uniform discretization ($\lfloor\ \rfloor$ denotes the floor function)

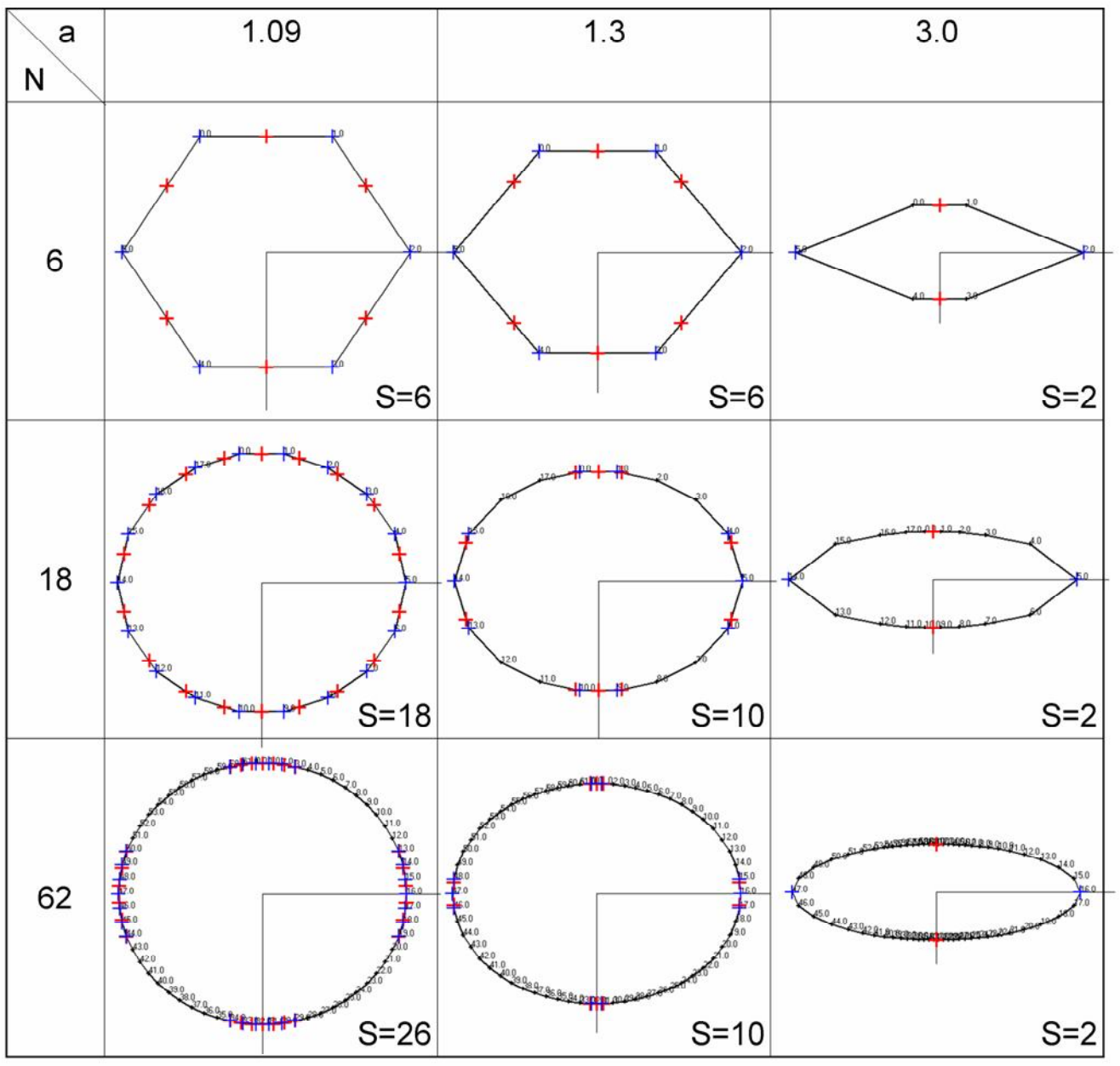

**Figure 4**: Uniform discretization of ellipse. Rows correspond to k=constant, i.e N=4k-2 constant number of discretizations points, columns correspond to constant axis ratio. S denotes the total number of stable micro-equilibria. Based on Table 1, in the N →∞ limit converges to $8\left\lfloor \frac{a^2}{2(a^2-1)} \right\rfloor + 2$

As we can observe, flocks are well-defined for elongated curves/surfaces and less prominent for nearly spherical geometries.

While local (micro) equilibria and flocks (macro equilibria) differ fundamentally, they both can be classified according to stability as defined above. Pebbles and rocks can only be physically at rest on a stable micro-equilibrium. However, the latter may belong either to a stable or to an unstable flock, the latter, ('mixed') scenario we call a rocking equilibrium. This concept is illustrated in Figure 5 for the simplest, planar case. "Rocking" motion corresponds to transitions between stable micro-equilibria inside the same unstable flock. We use the concept of rocking equilibrium to explain the behavior of rocking stones.

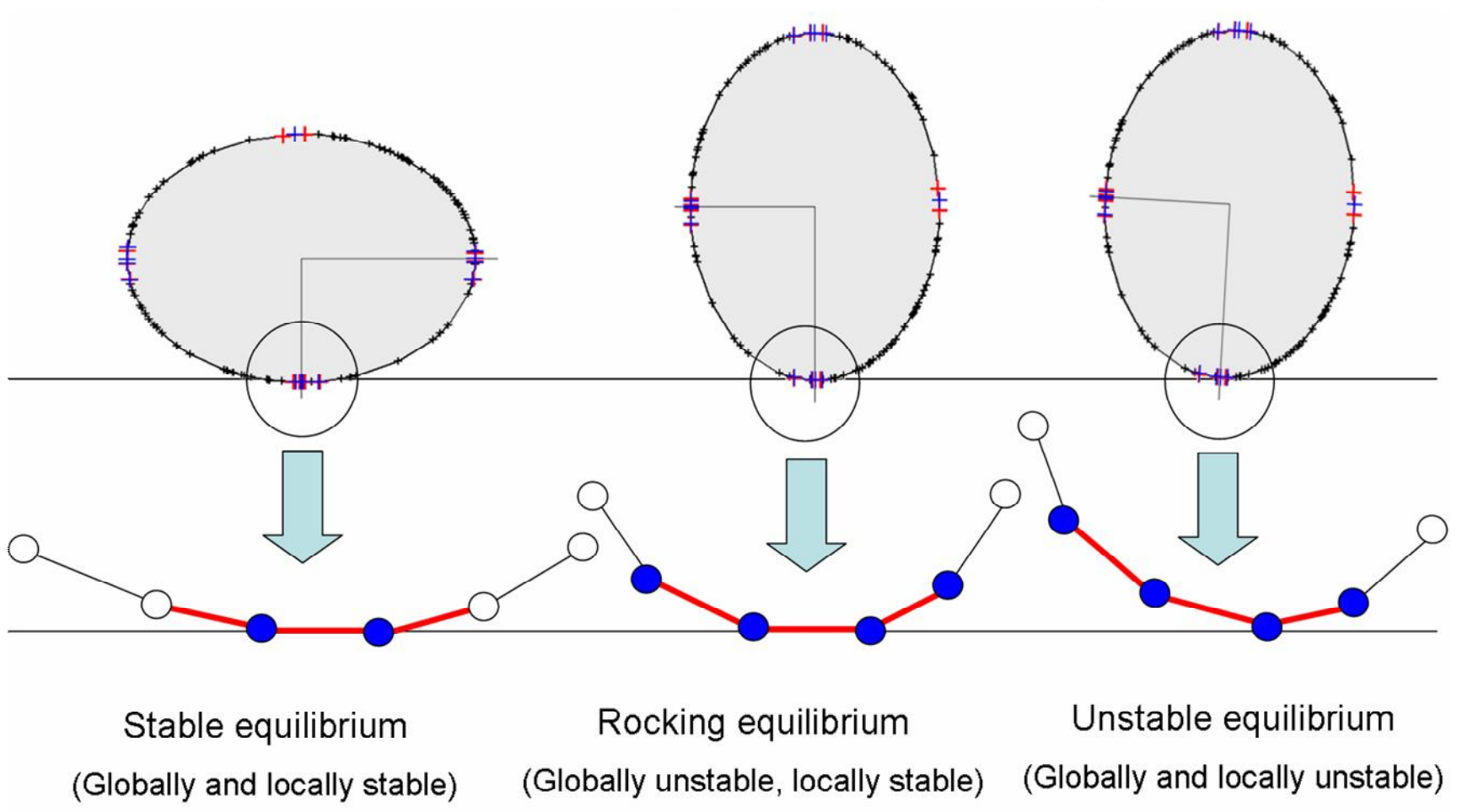

**Figure 5**: Stable, 'rocking' and unstable equilibrium for the planar, discretized ellipse.

## 1.2. Research goals, paper outline

Our primary goal is to show that the above illustrated scenarios are characteristic to real pebbles and rocks. To this end, we produced high-accuracy (0.1mm) 3D laser scans of 100 pebbles, and constructed their convex hulls carrying micro-equilibrium points (Section 2.1). Subsequently (Section 2.2), we identify these points and in Subsection 2.2.2. we construct the heteroclinic graph (i.e. the corresponding Morse-Smale complex) defining their topology. By systematic reduction of the complex (Subsection 2.2.3) we reduce the graph and identify the two mentioned scales (micro and macro) for equilibrium points. As we show in Section 3, our experiments reveal that these scales are well separated. Based on the scanned and computed results we identify one *rocking equilibrium* on a scanned pebble and produced it by physical experiment.

Our other goal is to validate the new classification for pebble shapes, suggested in our earlier paper (Domokos et al 2010). Here we proposed that instead of measuring lengths (serving the Zingg classification (Zingg 1935, Illenberger 1991), the experimenter should count the number and type of static equilibrium points. This method relies on simple hand experiments and the reliability of the latter is rooted in the separation of the micro and macro scales of equilibria. Our results show that these scales are well separated, thus the experimenter is able to reliably count macro-equilibria (flocks). Parameter $\nu$, introduced in the systematic reduction of the Morse-Smale complex can be also interpreted as the (scalar) measure of the reliability of the experimenting person. In Section 3 we show that comparison between our computer experiments and earlier hand experiments reveal not only that the method is consistent (i.e. one value of $\nu$ is characteristic for one experimenting person), the obtained data also reveals a close connection between flatness and the average number of equilibrium points.

## 2. Computer experiments

### 2.1. Scanning pebbles and the construction of their convex hull

In recent years, there has been a considerable interest in application of 3D laser scanning technology in geomorphological studies (Bourke et al 2008, Heritage and Hetherington 2007 MacMillan et al 2003, Nagihara et al 2004), since this new technology provides detailed information about the surface morphology of the scanned object. In our study we produced high-precision (0.1 mm) 3D laser scans of 5×20 pebbles, detailed information on the pebble samples is shown in Table 2 (Appendix). Pebbles were scanned using a Kondia NCT CNC milling machine with added SCANTECH 3D laser scanner (Fig. 6). With this equipment a 0.1 mm scanning accuracy can be achieved. We scanned pebbles in 5-7 different positions and we fitted together the resulting, partly overlapping point cloud pieces (Fig. 7A-C) by using RapidForm XOR reverse engineering software. Typical scans resulted in approx. 80000-100000 points per pebble, the joined point cloud was converted into a triangulated polyhedron with approx. 15000 faces, 10000 vertices and 25000 edges by using RapidForm (Fig. 7D). The resulting closed polyhedral surface was imported into our own code using the CGAL geometrical libraries (CGAL). Finally, we used an incremental approach in CGAL to build the convex hull which carries micro-equilibrium points (Fig. 7E).

A, 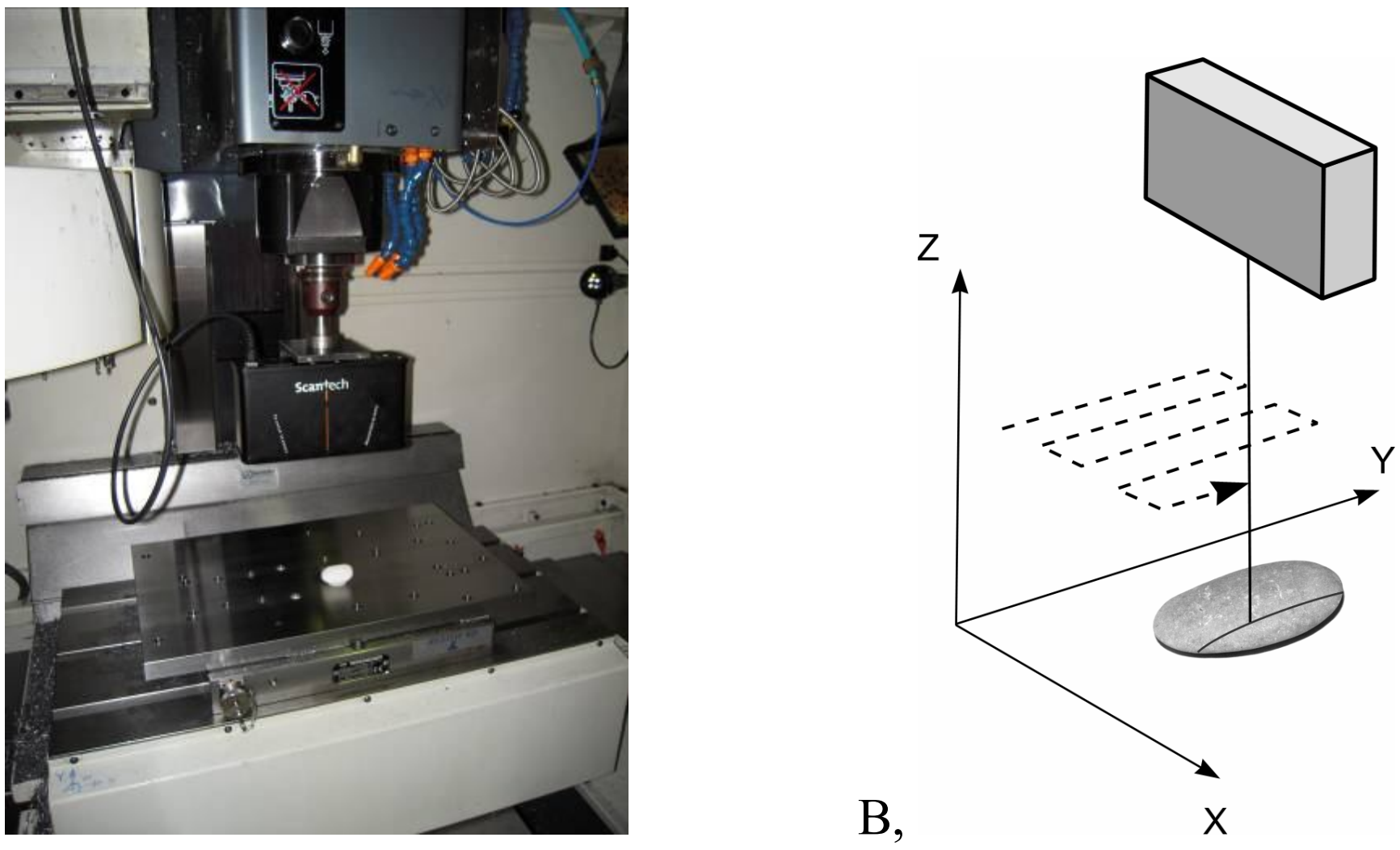 B,

**Figure 6**: a) Kondia NCT CNC machine with added SCANTECH 3D laser scanner b) Path of laser head above scanned pebble

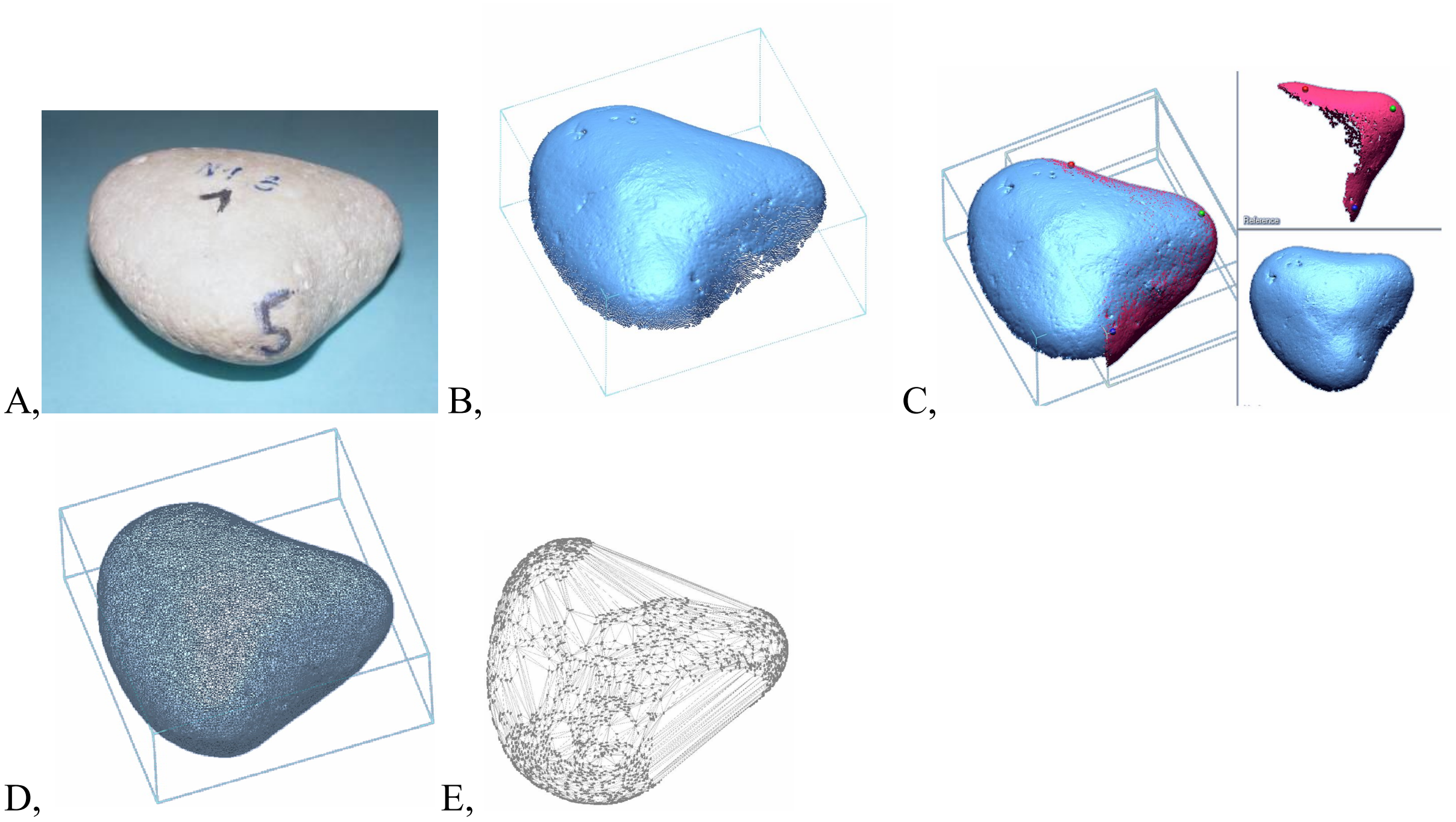

**Figure 7**: Illustration of the scanning process and the construction of the convex hull on a pebble (Pebble #1) A, picture of pebble B, one of the point cloud pieces C, joining point cloud pieces together D, point cloud converted into a triangulated polyhedron E, constructing the convex hull

## 2.2 Identification of equilibria, construction and simplification of the Morse-Smale Complex

### *2.2.1 Basic concepts*

Here we introduce the basic mathematical concepts, in particular equilibrium points and the corresponding *Morse-Smale complex* based on (Domokos et al 1994, Edelsbrunner et al 2003). We will rely on the systematic simplification of the Morse-Smale complex to identify flocks and equilibria on two separate scales.

By definition, at equilibrium positions the center of gravity $G$ of the pebble is located on the vertical line originating at the point $P$ of contact. As (Domokos et al 1994) showed for the 2D case, these points are identical to the singularities of the *gradient flow* associated with the scalar gravitational field, defined by the distance $R(\varphi)$. Our investigation is limited to *generic* shapes, i.e. we assume that all of the critical points of $R$ are non-degenerate thus $R$ is a *Morse function.*

In 3D, the gradient flow determined by the the scalar field $R(\varphi, \theta)$ can have 3 types of generic singularities: minima, maxima and saddles, corresponding to three generic types of equilibria to which we will briefly refer to as stable, unstable and saddle. On smooth surfaces all non-critical points belong to a unique route (an *integral line*) which is tangent to the gradient flow at each point and connects two critical points (Fig. 8A and B). Based on these integral lines the surface can be decomposed into *stable* and *unstable manifolds* associated with the aforementioned critical points. A stable manifold of a critical point contains all points where from the gradient flow reaches the critical point (along integral lines) and unstable manifolds contain all points which belong to integral lines having the critical point as origin. Stable and unstable manifolds can have various dimensions. On generic, smooth surfaces stable and unstable manifolds associated with saddles are 1-dimensional (lines) whereas stable and unstable manifolds of generic maxima and minima have even dimension (0 or 2).

A Morse function is called a *Morse-Smale function* if the stable and unstable manifolds intersect only transversally. In case of 2D surfaces, the transversally intersecting 1D manifolds (corresponding to saddles) define sub-domains which we call the *Morse-Smale cells,* the collection of which is referred to as the *Morse-Smale complex*. The interior of the cells corresponds to 2D (stable or unstable) manifolds and the vertices to 0-dimensional manifolds (critical points).

We will use a representation of the Morse-Smale complex by an embedded graph, with vertices corresponding to critical points ($S, U, H$ denoting the number of minima, maxima and saddles, respectively), edges corresponding to integral lines connecting saddles to maxima and minima (Fig. 8B and C). We briefly refer to this graph as the Morse-Smale graph of the surface. According to the Poincaré-Hopf Theorem (Arnold 1998), on a Morse-Smale graph corresponding to a convex pebble surface we always have

$$S+U-H=2 \tag{1}$$

We remark that Morse-Smale complexes and their simplification play an important role in structural biology in the description of molecular docking (Cazals et al 2003, Natarajan et al 2006, Agarwal et al 2006). It is also used in oceanography the extremes of temperature and the most practical applications belong to imaging such as MRI in medicine and X-ray crystallography in molecular biology (Edelsbrunner et al 2003:2).

Since we model the pebbles by convex polyhedrons arising from our 3D scans (cf. Section 2.1), so the Morse-Smale theory for *piecewise linear manifolds* (Forman 1998) has to be applied. While this theory also predicts the same three types of generic singularities (maxima, minima and saddles), equation (1) is applicable and the dimensions of the associated stable and unstable manifolds also remain unchanged, some of the non-critical points (typically points along edges) can belong to more routes between critical points, i.e. the integral lines are not uniquely defined (Fig. 8D-F).

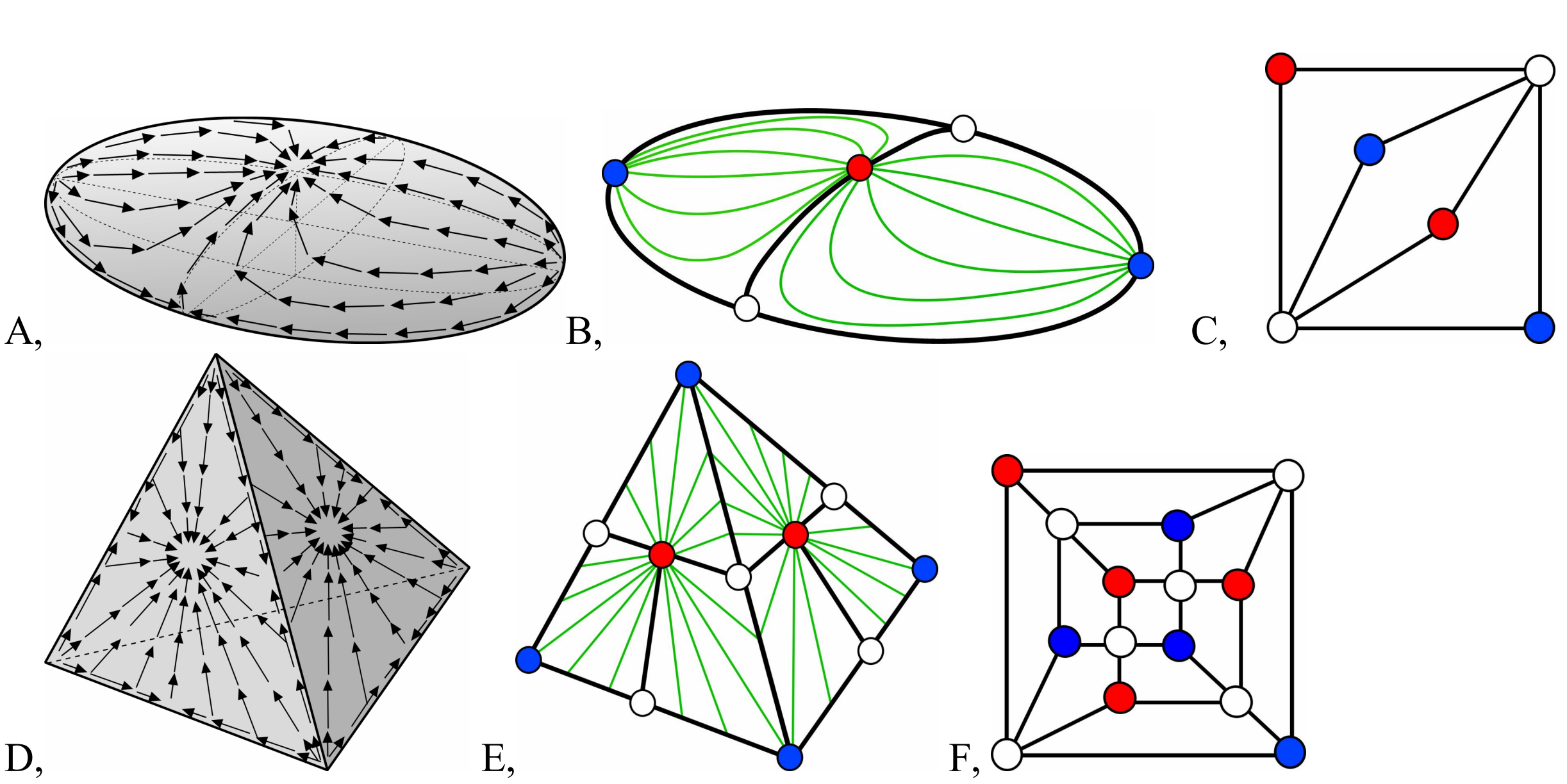

**Figure 8**: A, gradient flow on a smooth surface (ellipsoid) B, integral lines (green) and Morse-Smale graph (bold black lines) on the surface of a smooth ellipsoid. Red/ blue/ white vertices denote stable/ unstable/ saddle type equilibria. C, Morse-Smale graph of the ellipsoid on the plane D, gradient flow on a polyhedron (tetrahedron) E, some routes between critical points (green) and Morse-Smale graph (bold black lines) on the surface of a tetrahedron F, Morse-Smale graph of the tetrahedron on the plane. Observe that for polyhedra, non-critical points along edges belong to more routes between critical points.

### *2.2.2 Construction of the Morse-Smale graph*

As indicated in Section 2.1, the scanned pebbles are ultimately represented by convex polyhedra. In the first step we identify the equilibrium points on the polyhedral surface then we trace the integral lines connecting saddles to maxima and minima.

Due to the convexity of the polyhedron, identifying the *critical points* of $R(\varphi,\theta)$ is simple:

1. Stable points are located on facets,
2. Saddle points are located on edges,
3. Unstable points are located on vertices of the polyhedron.

Let G denote the center of gravity of the polyhedron. For having a stable point $P_S$ on a facet F the normal of F through $P_S$ must contain G as well. For having a saddle point $P_H$ on an edge E two conditions must be fulfilled: a) the normal plane of E at $P_H$ must contain G and b) the normal plane to $GP_H$ at $P_H$ cannot intersect the polyhedron except the line E. For having an unstable point $P_U$ at a vertex V the normal plane to GV at V cannot intersect the polyhedron except at point V itself.

The next step is identifying the edges of the graph in the complex, i.e. to trace the integral lines connecting saddles to maxima and minima. Due to our assumption about genericity, we do not have to handle degenerate cases, i.e. a saddle point in our complex is always the origin of 4 edges, two of which lead to stable points, the other two originate at unstable points. The integral curves leading to and originating from the saddle can be traced by starting at the saddle and following the gradient of *R* forward or backward until we hit the adjacent critical point (maximum or minimum). This strategy – which in fact is a *steepest ascent* approach – defines the adjacent critical points uniquely. After identifying the paths to and from all saddles we obtain the full Morse-Smale graph.

### *2.2.3 Simplification of the Morse-Smale graph*

Separating scales and the identification of flocks can be done via the systematic simplification of the Morse-Smale graph. As stated in the introduction, a flock corresponds to a localized collection of micro-equilibria, all located in a potential well (stable flock) or rather, on top of a potential hill or in the vicinity of a saddle (unstable flock). The goal of the simplification is to remove these micro-equilibria so that only the dominant, global equilibria persist, corresponding to the large-scale variation of the potential landscape.

We perform the simplification on the Morse-Smale graph (rather than on the surface itself). (It is worth mentioning that according to the algorithm in (Bremer at al 2003) the modification of the surface geometry according to the simplification is also possible, however, it is not of interest from our current point of view.) Based on (Edelsbrunner et al 2003), we use *cancellations* to simplify the Morse-Smale graph (Fig. 9). Each cancellation can be seen as a double *edge contraction* that removes two neighboring vertices from the graph (an edge contraction is an operation which removes an edge from a graph while simultaneously merging together the two vertices it previously connected). We cancel critical points in pairs, i.e. we remove a saddle point and an unstable point, or a saddle point and a stable point in each step, so we keep consistence with Equation (1).

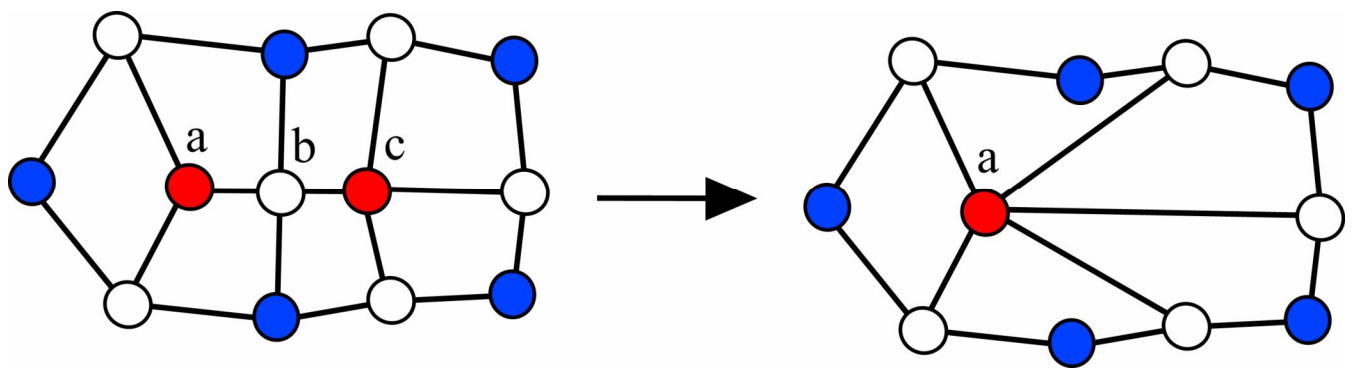

**Figure 9**: Cancellation that removes equilibrium points b and c from the Morse-Smale graph

To determine the order of the cancellations we assign a 'weight' to each edge in the graph. The weight of an edge is the difference in the value of the function $R(\varphi,\theta)$ between the ends of the edge i.e. the potential difference between the two, adjacent critical points. Formally, we assign a value $\nu_{ij}$ to each edge connecting equilibrium points $i$ and $j$ by the following definition:

$$\nu_{ij} = \frac{\left|R(\varphi_i,\theta_i) - R(\varphi_j,\theta_j)\right|}{R_{\max}},$$

where $R_{\max}$ is the distance of the furthermost vertex of the surface from center of gravity G. Observe, that $0 \le \nu_{ij} \le \nu_{\max} \le 1$, where $\nu_{\max} = \frac{\left|R_{\max} - R_{\min}\right|}{R_{\max}}$ .

As mentioned, a flock is located in a potential well (or on a potential hill), so potential differences *inside* a flock are small compared to the large-scale changes in $R(\varphi,\theta)$. Thus, if we rank cancellations according to increasing value of $\nu_{ij}$ (i.e. increasing potential differences), we can remove micro-equilibria from the flocks as long as there remain only the macro-equilibrium points in the graph, which are associated with the dominant changes of the potential function $R(\varphi,\theta)$.

By this process we define a one-parameter family of graphs $G(\nu)$. $G(0)$ is identical with the original graph and $G(\nu)$ is the graph produced from $G(0)$ by *cancellations* of all edges with $\nu_{ij}<\nu$. It is easy to see that $G(\nu_{max})$ will only have one stable and one unstable equilibrium point. This family of graphs defines the scalar function $N(\nu)$ as the total number of equilibrium points of $G(\nu)$ and its normalized form $n(\nu)=N(\nu)/N(0)$. The function $n(\nu)$ (which can be plotted on the unit square) plays a fundamental role in our work which we describe below.

This cancellation strategy does not reveal flocks directly, however, by plotting $n(\nu)$ versus $\nu$ we obtain a simple plot on which flocks can be identified. Observe that $n(\nu)$ is monotonically decreasing since, with increasing $\nu$, points are always *removed* from the Morse-Smale graph. If $n(\nu)$ decreases gradually, in small steps, then one can not identify flocks. However, a sudden jump followed by a plateau in $n(\nu)$ reveals that equilibria exist on two separate scales since for a broad range of $\nu$ the function $n(\nu)$ remains constant and this constant corresponds to the number of macro-equilibria. The initial, high values of $n(\nu)$ correspond to micro-equilibria (local scale). In the next section we will compute $n(\nu)$ for scanned pebbles and discuss the resulting plots.

The above described simplification algorithm is not only suitable to identify equilibria on two (micro and macro) scales, parameter $\nu$ can also be used to model our hand experiments proposed in our previous paper (Domokos et al 2010). In that paper we showed that the number and type of macro-equilibrium points describe the geometry of a pebble well, therefore we proposed to classify pebble shapes based on simple hand experiments in which we count the macro-equilibrium points (flocks) of the pebble. The reliability of the hand experiments depends on the separation of micro and macro scales i.e. whether the experimenter can identify flocks or not. Parameter $\nu$ can be considered as the scalar measure of the inaccuracy of the experimenting person: below this threshold of the potential difference, the experimenter misses to distinguish micro-equilibrium points and substitutes them with a single equilibrium point. In Section 3 we will validate our hand experiments by comparing them to the computer experiments in which we model the reliability of the experimenter with the parameter $\nu$.

## 3. Interpretation of the results: identifying rocking equilibria

The primary goal of this paper is to explain the mechanics of rocking stones with the concept of rocking equilibrium. Therefore, we identified and physically produced a rocking equilibrium on a scanned pebble.

Based on Section 2.2, we identified micro-equilibrium points and constructed the original Morse-Smale graph, G(0) on a typical pebble (Fig. 10A). Figure 10B shows the systematic reduction of G(0), followed by the logarithmic plot of $n(\nu)$. Observe the long plateau, indicating the well-separated macro scale of equilibrium points. Figure 10C shows macro-equilibria on the pebble's surface and the corresponding, reduced Morse-Smale graph.

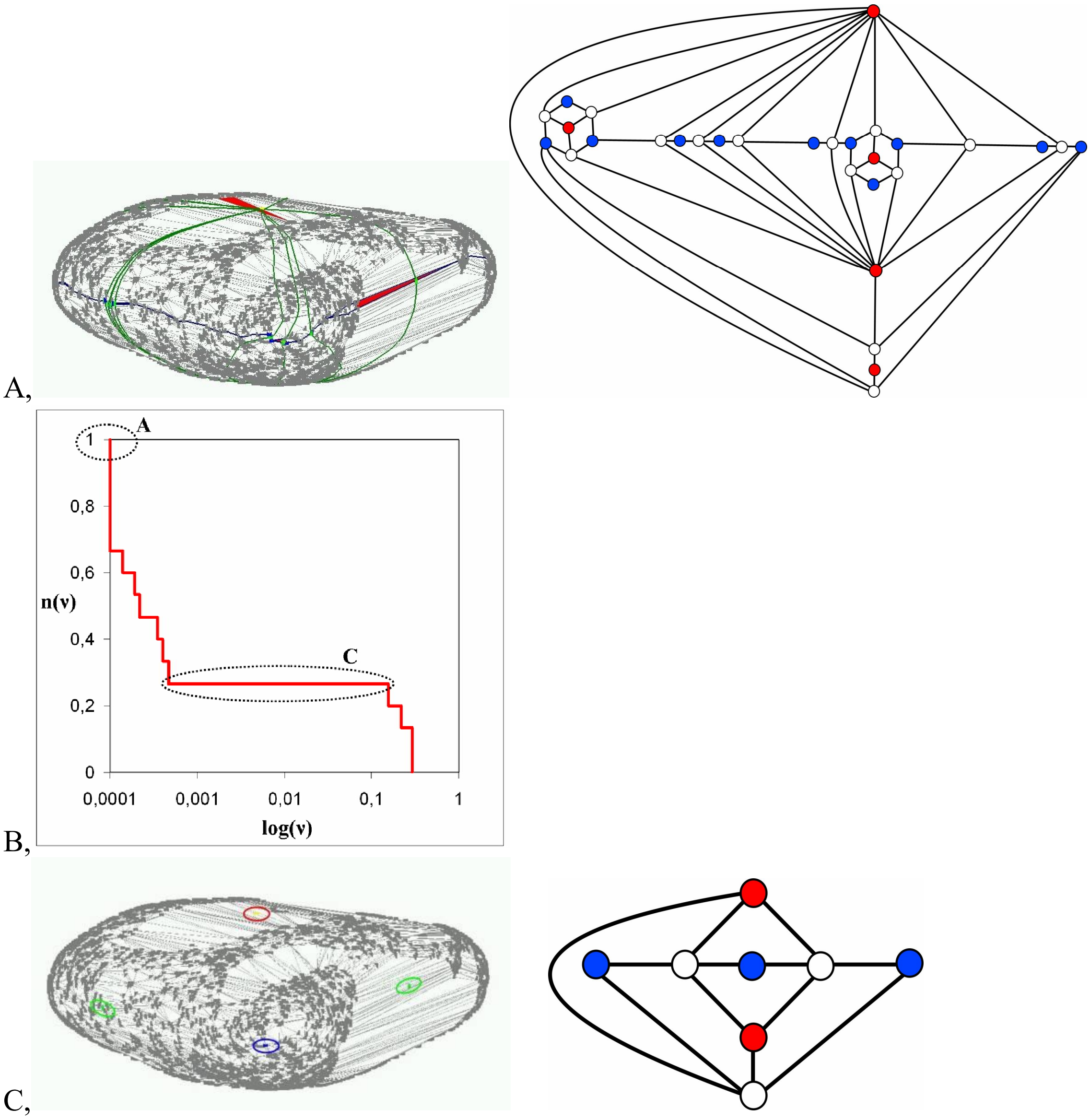

**Figure 10**: separation of scales on Pebble #1:
A, micro-equilibria on the pebble's surface and the original Morse-Smale graph, G(0) on the pebble and in the plane B, log(ν) – n(ν) diagram. The diagram has a sudden jump at ν=0 and a long plateau after that, indicating that micro and macro scales are well separated C, stable and unstable flocks (macro-equilibria) on the pebble's surface and the corresponding reduced Morse-Smale graph in the plane

On Figure 11 we marked two different micro-equilibrium points on the pebble's surface: a stable micro-equilibrium in a stable flock and a stable micro-equilibrium in an unstable flock i.e. a rocking equilibrium. Usually, the pebble is balanced on the former, however, it is possible to balance the pebble in a counter-intuitive position corresponding to the marked rocking equilibrium as well (photos of Fig. 11). This equilibrium is very sensitive, as it belongs to an unstable flock: even a small force causes to dislodge the pebble, not returning to the rocking equilibrium again. However, in case of a large and heavy stone, human force usually causes the stone only to rock as long as it finally returns to the rocking equilibrium (or to an other stable micro-equilibria in the same unstable flock). This experiment validates the concept of rocking equilibrium which, in turn, explains the existence of rocking stones.

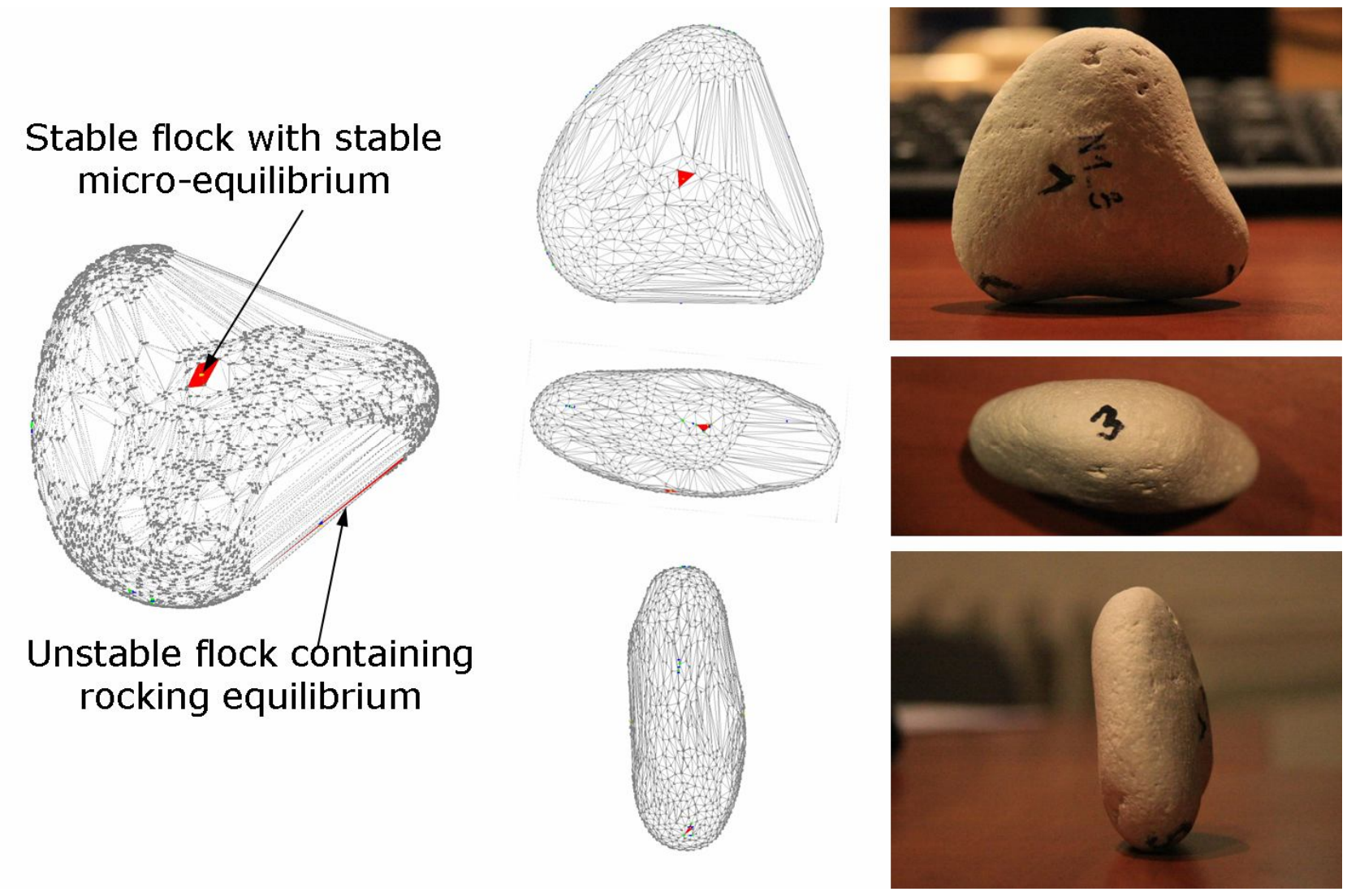

**Figure 11**: identifying and physically producing a rocking equilibrium on Pebble #1.

Our other goal is to verify the accuracy of the hand experiments on which our new classification system for pebble shapes (Domokos et al 2010) is based. The sudden jump of the $\nu - n(\nu)$ diagram near $\nu=0$ and the long plateau indicates that the micro and macro scales are well separated, thus macro-equilibria (flocks) can be reliably counted in a hand experiment i.e. our new classification system is practically applicable as long as pebbles are not too close to the sphere.

Computer experiments can also be compared to our earlier hand experiments (Szabó and Domokos 2011). In those hand experiments we collected 22 pebble samples, each sample containing 50 pebbles. (Detailed data is shown in Table 2, see Appendix. Here we also compare the results of hand experiments to computer experiments). We counted macro-equilibrium points on the pebbles and measured axis lengths $a>b>c$ proposed in the Zingg system (Zingg 1935). Then, we computed the average axis ratios ($c/b$ and $c/a$), the average number of equilibrium points (E(S), E(U)) and the standard deviation of equilibrium points ($\sigma$(S), $\sigma$(U)) for each pebble sample. Hand experiments show a close relationship between the average number of stable equilibrium points (E(S)) and the average flatness (axis ratios $c/b$ and $c/a$) of the pebbles (Fig. 12A, purple and yellow trend lines), indicating that equilibrium classification is closely connected to the Zingg system. This result also validates that our new classification system is suitable to describe pebble shapes. We also find an interesting relationship between the average number (E(S),E(U)) and the standard deviation ($\sigma$(S), $\sigma$(U)) of equilibrium points (Fig. 12B).

As mentioned, parameter $\nu$ is the measure of the inaccuracy of the experimenting person: $\nu=0$ stands for an experimenter with absolute accuracy who recognizes each micro-equilibrium point, increasing value of $\nu$ models the effect of increasing human error in a hand experiment. We examined whether our earlier hand experiments can be modeled with this single parameter, so we computed the Morse-Smale graphs for the $5\times20$ scanned pebbles (cf. the first 5 rows of the Appendix) and applied the simplification algorithm for them. For any given value of $\nu$ E(S), E(U), $\sigma$(S) and $\sigma$(U) can be computed for the 5 scanned samples and the values can be compared to the result of hand experiments. We found that there is a certain value of parameter $\nu$ ($\nu=0.0035$) at which data of computer experiments matches our hand experiments well for all investigated variables (Fig. 12A and B), i.e. parameter $\nu$ appears to characterize well the human experimenter. As our hand experiments were done by *one* person, this indicates that one experimenter counts equilibrium points always with the same accuracy i.e. hand experiments are consistent. This result validates the practical applicability of our new shape classification system based on static equilibrium points.

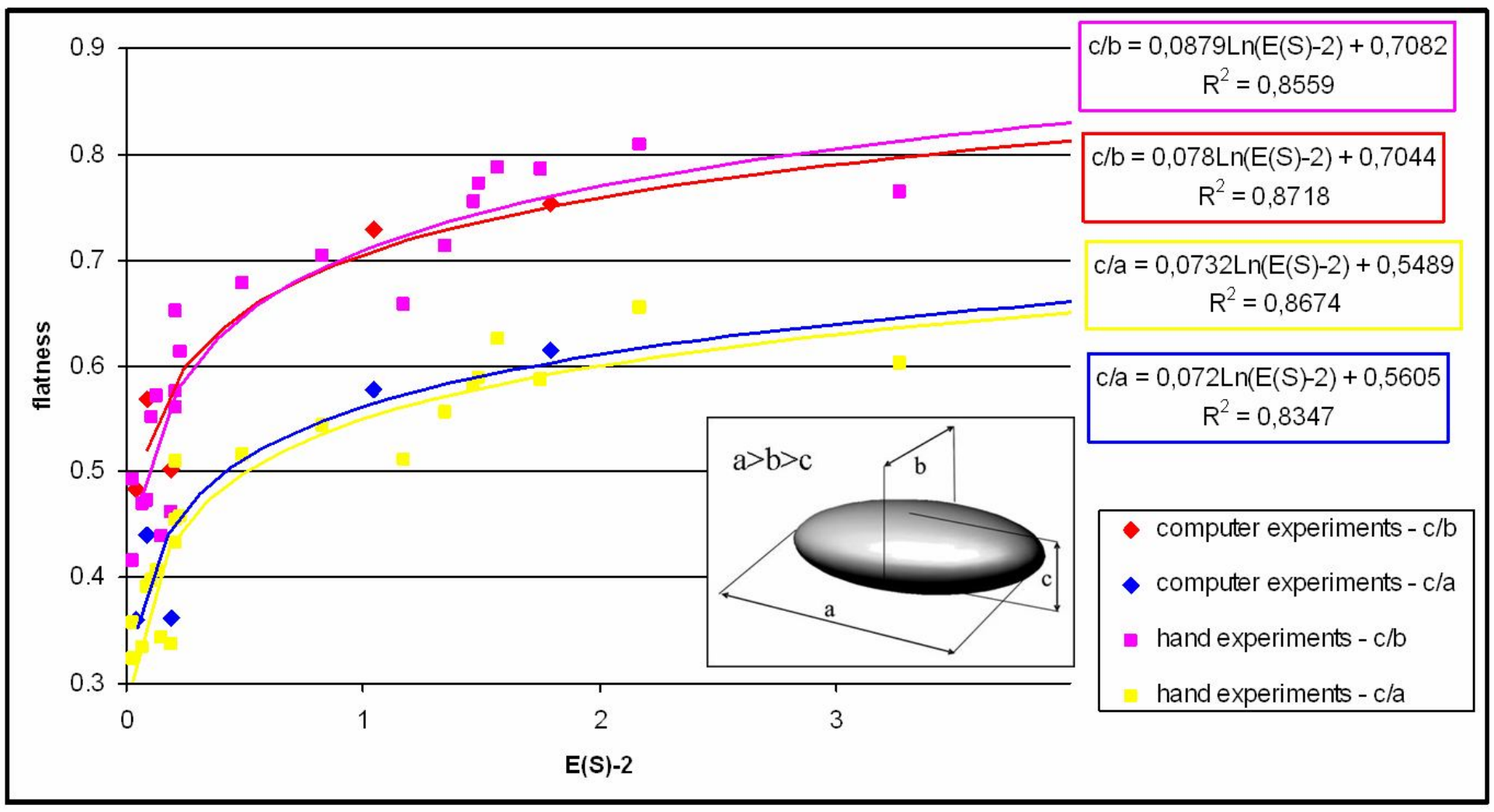

A,

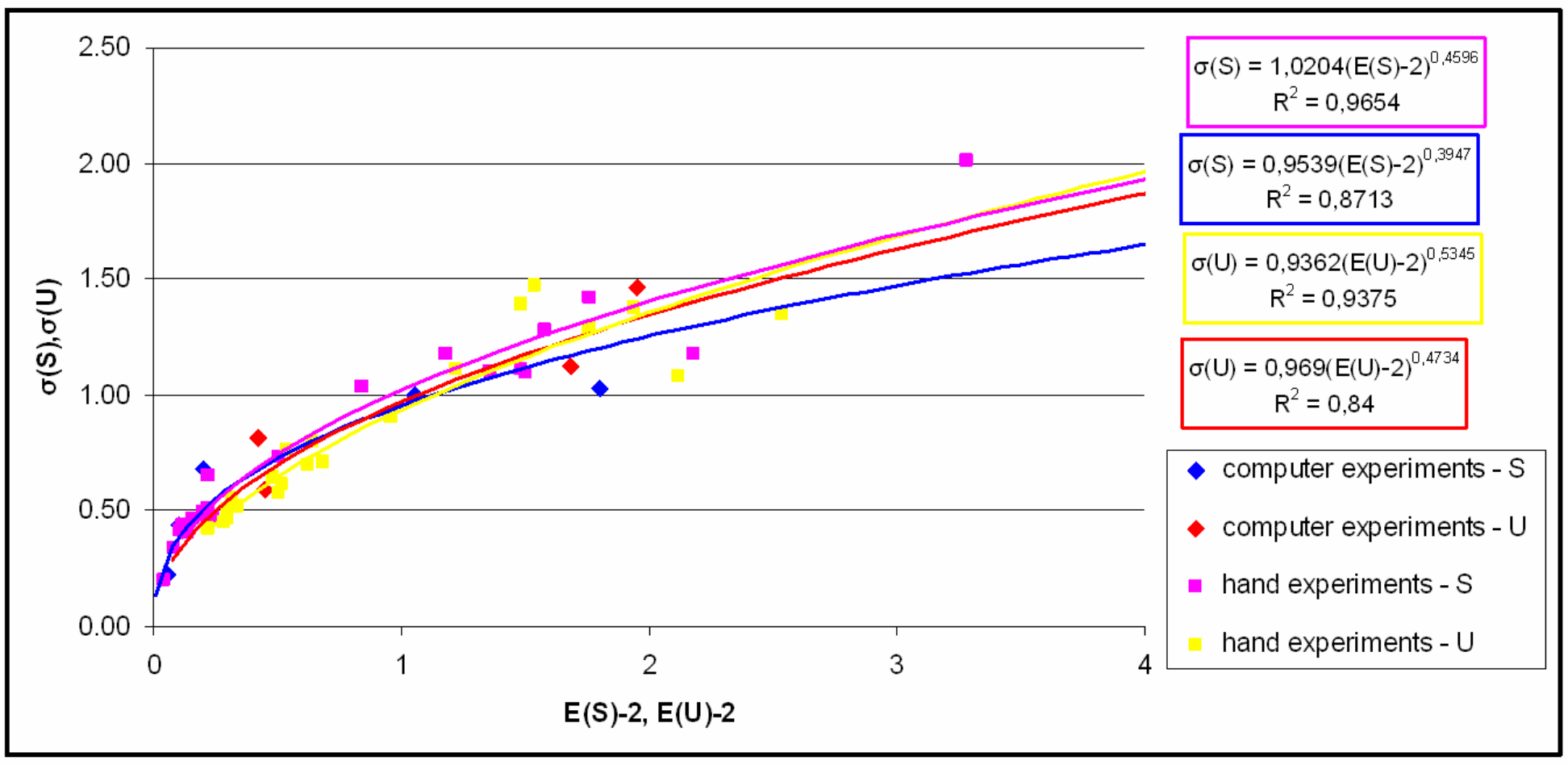

B,

**Figure 12**: comparison between hand experiments and computer experiments. Purple and yellow data points and trend lines represent hand experiments, red and blue data points and trend lines show the result of computer experiments at υ=0.0035. Computer experiments match hand experiments well.
A, relationship between the average number of stable equilibrium points (E(S)) and flatness (axis ratios c/b and c/a)
B, relationship between the average number (E(S),E(U)) and the standard deviation (σ(S), σ(U)) of equilibrium points

## 4. Summary

In this paper we showed that static equilibria of rocks and pebbles typically exist on two, well-separated scales. Local, micro-equilibria form flocks and the latter can be regarded as global, macro-equilibria. We demonstrated this phenomenon by analyzing the high-precision scans of real pebbles, identifying all micro-equilibria by computer. Subsequently, by a systematic simplification of the Morse-Smale complex, we pinpointed flocks of micro-equilibria forming global macro-equilibria.

This new concept is not only interesting because it can explain the phenomenon of rocking stones, it also provides a firm basis and justification for the reliability of field experiments in the classification of pebbles based on the number of their equilibrium points. Last but not least, this new concept may serve as an interesting illustration to a fundamental theorem in convex geometry.

## 4.1 Rocking equilibria

Rocking stones, balanced in counter-intuitive positions have always intrigued geologists. The concept of micro- and macro-equilibria provides a framework to define and explain this phenomenon. Both micro- and macro equilibrium points can be either stable or unstable and any rock can be only balanced on a stable micro-equilibrium. If the latter belongs to a stable flock (macro-equilibrium) then the stone is globally at a regular, stable position. If, however, the stable micro-equilibrium is part of an unstable flock, then we have a rocking equilibrium. To illustrate this concept, by using the above definition and high-precision 3D scanning, we identified one rocking equilibrium point on the surface of a pebble and managed to balance the pebble in the counter-intuitive position. Rocking stones are large-scale illustrations of the same concept.

## 4.2 Field experiments with pebbles

In an earlier study we proposed to classify pebbles based on the number and type of their equilibrium points. We showed that this classification carried more information than the classical Zingg system. The practical applicability of the new method is hinged on the reliability of hand experiments. The separation of micro- and macro scales implies that in hand experiments one can reliably count macro-equilibria. We verified this by analyzing 5x20 3D scans and comparing the results of computer experiments to hand experiments. We introduced a scalar parameter measuring human error and found that the method is consistent because for the same experimenter we measured parameter values in a narrow range. We found that the number of global, macro-equilibria remains constant for a broad range of this parameter, thus the result of the field experiments are robust and do not depend on the experimenting person.

## 4.3 Illustration of a mathematical result

Convex geometry deals with properties of typical convex bodies which are neither smooth nor polyhedral. Zamfirescu's Theorem (Zamfirescu 1995) states that a typical convex body has infinitely many equilibrium points. The theorem does not specify the location or distribution of these equilibria. The theorem does not apply directly to polyhedra nor to smooth manifolds, however, one hopes to find analogous phenomena on these surfaces. Our results show that in case of finely discretized convex surfaces (i.e. convex polyhedra with many faces) one can indeed find large numbers of local equilibrium points and this appears to be analogous to Zamfirescu's Theorem. Also, we found that these local micro-equilibria are strongly localized and these collections (flocks) can be interpreted as macro-equilibria. This provides an intuitive illustration to Zamfirescus's Theorem.

## 4.4 Outlook and further potential applications

As we showed, hand experiments can reliably determine the number of macro-equilibria, computer scans provide the number of micro-equilibria. While the latter are certainly more time-consuming, the resulting data is very useful to describe the detailed morphology of pebbles. Besides the actual number of equilibrium points, computer scans also detect the Morse-Smale complex associated with the surface and this graph can be also used to describe the morphology of the stone. The concept of micro- and macro-equilibria on separate scales can be also used to describe evolution of morphology in abrasion processes.

## Acknowledgement

This research was supported by OTKA grant T072146. Pebbles scans have been performed by Gábor Kovács, Zoltán Iváncsics, Mátyás Farkas and Csaba Haraszkó in the framework of a student project, the scanned point cloud pieces were fitted together by Zsófia Salát and Benedek Kiss. We thank Prof. Gyula Mátyási for his kind advice and help with the scans.

## References

Agarwal PK, Edelsbrunner H, Harer J, Wang Y (2006) Extreme elevation on a 2-manifold. Discrete Comput Geom 36:553–572

Anooshehpoor A, Brune JN, Zeng Y (2004) Methodology for obtaining constraints on ground motion from precariously balanced rocks. Bull Seismol Soc Am 94:285-303

Arnold VI (1998) Ordinary differential equations, 10th Printing, Mit Press, Cambridge

Bell JW, Brune JN, Liu T, Zreda M, Yount JC (1998) Dating precariously balanced rocks in seismically active parts of California and Nevada. Geology 26:495-498

Bourke M, Viles H, Nicoli J, Lyew-Ayee P, Ghent R, Holmlund J (2008) Innovative applications of laser scanning and rapid prototype printing to rock breakdown experiments. Earth Surf Process Landforms 33:1614–1621

Bremer PT, Edelsbrunner H, Hamann B, Pascucci V (2003) A multiresolution data structure for two-dimensional Morse-Smale functions. Proceedings of the 14th IEEE Visualization Conference VIS 2003

Brune JN, Anooshehpoor A, Purvance MD, Brune RJ (2006) Band of precariously balanced rocks between the Elsinore and San Jacinto, California, fault zones: Constraints on ground motion for large earthquakes. Geology 34:137-140

Cazals F, Chazal F, Lewiner T (2003) Molecular shape analysis based upon the Morse-Smale complex and the Connoly function. Proceedings of the 19th ACM Symposium on Computational Geometry, Socg'03, San Diego, California

CGAL, Computational Geometry Algorithms Library, http://www.cgal.org

Domokos G, Papadopoulos J, Ruina A (1994) Static equilibria of rigid bodies: is there anything new? J Elasticity 36:59-66

Domokos G, Sipos A, Szabó T, Várkonyi P (2010) Pebbles, shapes and equilibria. Math Geosciences 42:29-47

Edelsbrunner H, Harer J, Zomorodian A, (2003) Hierarchical Morse-Smale complexes for piecwise linear 2-manifolds. Discrete Comput Geom 30:87–107

Edelsbrunner H, Harer J, Natarajan A, Pascucci V, (2003) Hierarchical Morse-Smale complexes for piecwise linear 3-manifolds. Proceedings of the 19th ACM Symposium on Computational Geometry, Socg'03, San Diego, California

Forman R, (1998) Morse Theory for cell complexes. Adv Math 134:90-145

Heritage GL, Hetherington D (2007) Towards a protocol for laser scanning in fluvial geomorphology. Earth Surf Process Landforms 32:66–74

Illenberger W (1991) Pebble shape (and size!) J of Sedimen Res 61:756

Linton DE (1955) The problem of tors. Geogr J 121:470-481+487

Macmillan RA, Martin TC, Earle TJ, Mcnabb DH. (2003) Automated analysis and classification of landforms using high-resolution digital elevation data: applications and issues. Can J Rem Sens 29:592–606

Nagihara S, Mulligan KR, Xiong W (2004) Use of a three-dimensional laser scanner to digitally capture the topography of sand dunes in high spatial resolution. Earth Surf Process Landforms 29:391–398

Natarajan V, Wang Y, Bremer PT, Pascucci V, Hamann B (2006) Segmenting molecular surfaces. Comput Aided Geom D, 23:495–509

Szabó T, Domokos G (2011) A new classification system for pebble and crystal shapes based on static equilibrium points. Central European Geology (accepted for publication)

Wikipedia: Rock Balancing, http://en.wikipedia.org/wiki/Rock_balancing

Wikipedia: Rocking Stones, http://en.wikipedia.org/wiki/Rocking_Stones

Zamfirescu T (1995) How do convex bodies sit? Mathematica 42:179-181

Zingg T (1935) Beitrag zur Schotteranalyse. Schweizer Miner Petrog Mitt, 15:39-140

# Appendix

| # ID | Location | Sample size: Hand experiments / Computer experiments | Characteristic rock type | Average pebble size (axis *b* in mm) | E(S) | E(U) | σ(S) | σ(U) | c/b | c/a |
|---|---|---|---|---|---|---|---|---|---|---|
| | | | | | Hand experiments / Computer experiments at υ=0.0035 | | | | | |
| 1. T1 | Marathonis beach, Zakinthos, Greece | 50 / 20 | limestone marlstone | 29.3 | 3.36 / 3.05 | 4.12 / 3.68 | 1.10 / 1.00 | 1.08 / 1.13 | 0.71 | 0.56 |
| 2. A1 | Argassi beach, Zakinthos, Greece | 50 / 20 | sandstone limestone | 18.8 | 3.48 / 3.80 | 3.48 / 3.95 | 1.11 / 1.03 | 1.39 / 1.47 | 0.75 | 0.58 |
| 3. K1 | Kamari beach, Santorini, Greece | 50 / 20 | andesite quartzite | 30.8 | 2.22 / 2.10 | 2.52 / 2.45 | 0.65 / 0.44 | 0.61 / 0.59 | 0.56 | 0.45 |
| 4. K3 | Kamari beach, Santorini, Greece | 50 / 20 | andesite quartzite | 17.9 | 2.08 / 2.20 | 2.28 / 2.45 | 0.34 / 0.68 | 0.45 / 0.59 | 0.47 | 0.33 |
| 5. K4 | Kamari beach, Santorini, Greece | 50 / 20 | andesite quartzite | 12.0 | 2.04 / 2.05 | 2.22 / 2.42 | 0.20 / 0.22 | 0.42 / 0.82 | 0.49 | 0.36 |
| 6. S1 | Via Mala, Graubünden, Switzerland | 50 | mica schist quartz schist | 33.9 | 2.16 | 2.64 | 0.47 | 0.80 | 0.44 | 0.34 |
| 7. S2 | Via Mala, Graubünden, Switzerland | 50 | mica schist quartz schist | 31.0 | 2.20 | 2.28 | 0.49 | 0.45 | 0.46 | 0.34 |
| 8. S3 | Via Mala, Graubünden, Switzerland | 50 | mica schist quartz schist | 29.9 | 2.04 | 2.62 | 0.20 | 0.70 | 0.42 | 0,32 |
| 9. T2 | Marathonis beach, Zakinthos, Greece | 50 | limestone marlstone | 24.9 | 3.50 | 3.54 | 1.09 | 1.47 | 0.77 | 0.59 |
| 10. A2 | Argassi beach, Zakinthos, Greece | 50 | andesite tuff sandstone | 39.5 | 3.18 | 3.76 | 1.17 | 1.29 | 0.66 | 0.51 |
| 11. N1 | Navagio beach, Zakinthos, Greece | 50 | limestone | 37.2 | 2.22 | 2.48 | 0.51 | 0.65 | 0.58 | 0.43 |
| 12. V1 | Danube bank, Vác, Hungary | 50 | quartzite | 28.3 | 2.12 | 2.30 | 0.44 | 0.46 | 0.55 | 0.40 |
| 13. M1 | Monolithos beach, Santorini, Greece | 50 | basalt andesite | 37.4 | 2.10 | 2.50 | 0.42 | 0.58 | 0.47 | 0.39 |
| 14. M2 | Monolithos beach, Santorini, Greece | 50 | pumice | 28.1 | 2.84 | 2.68 | 1.04 | 0.71 | 0.70 | 0.54 |
| 15. R1 | Red Beach, Santorini, Greece | 50 | andesite | 26.7 | 2.14 | 2.32 | 0.40 | 0.55 | 0.57 | 0.41 |
| 16. R2 | Red Beach, Santorini, Greece | 50 | andesite | 25.3 | 2.24 | 2.34 | 0.48 | 0.52 | 0.61 | 0.46 |
| 17. R3 | Red Beach, Santorini, Greece | 50 | andesite | 28.6 | 2.22 | 2.30 | 0.46 | 0.51 | 0.65 | 0.51 |
| 18. R4 | Akrotiri beach, Santorini, Greece | 50 | pumice | 24.9 | 3.58 | 2.96 | 1.28 | 0.90 | 0.79 | 0.63 |
| 19. K2 | Kamari beach, Santorini, Greece | 50 | pumice | 24.5 | 4.18 | 3.94 | 1.17 | 1.38 | 0.81 | 0.66 |
| 20. P1 | Perissa Beach, Santorini, Greece | 50 | pumice | 15.9 | 2.50 | 2.54 | 0.74 | 0.76 | 0.68 | 0.52 |
| 21. C1 | Janubio beach, Canary Islands, Spain | 50 | basalt | 30.7 | 3.76 | 3.22 | 1.42 | 1.11 | 0.79 | 0.59 |
| 22. C2 | Janubio beach, Canary Islands, Spain | 50 | basalt | 32.3 | 5.28 | 4.54 | 2.01 | 1.34 | 0.76 | 0.60 |

**Table 2:** detailed data of the collected 22 pebble samples and comparison between hand experiments and computer experiments. Values c/b and c/a are the average axis ratios used in the Zingg classification system, E(S) and E(U) are the average number of stable/unstable equilibrium points, σ(S) and σ(U) the standard deviation of stable/unstable equilibrium points, respectively. 20-20 pebbles from Sample 1-5 (shaded rows) were scanned and analyzed as described in Section 2. Results of computer experiments are given at υ=0.0035, where the investigated variables of computer experiments are the closest to that of hand experiments.